\documentclass[10pt,letterpaper,twocolumn]{article} 

\usepackage{ol2}
\usepackage{amsmath,amssymb,cite}
\usepackage[colorlinks,linkcolor=blue,citecolor=blue,urlcolor=blue,anchorcolor=blue]{hyperref}

\begin{document}

\twocolumn[ 

\title{Soliton pair generation in the interactions of Airy and nonlinear accelerating beams}


\author{Yiqi Zhang,$^{1,5}$ Milivoj Beli\'c,$^{2,6}$ Zhenkun Wu,$^1$ Huaibin Zheng,$^1$ Keqing Lu,$^3$ Yuanyuan Li,$^4$ Yanpeng Zhang$^{1,*}$}

\address{
$^1$Key Laboratory for Physical Electronics and Devices of the Ministry of Education \&
Shaanxi Key Lab of Information Photonic Technique,
Xi'an Jiaotong University, Xi'an 710049, China
\\
$^2$Science Program, Texas A\&M University at Qatar, P.O. Box 23874 Doha, Qatar \\
$^3$School of Information and Communication Engineering, Tianjin Polytechnic University, Tianjin 300160, China \\
$^4$Institute of Applied Physics, Xi'an University of Arts and Science, Xi'an 710065, China\\
$^5$zhangyiqi@mail.xjtu.edu.cn\\
$^6$milivoj.belic@qatar.tamu.edu\\
$^*$Corresponding author: ypzhang@mail.xjtu.edu.cn
}

\begin{abstract}
  We investigate numerically the interactions of two in-phase and out-of-phase
  Airy beams and nonlinear accelerating beams in Kerr
  and saturable nonlinear media, in one transverse dimension.
  We find that bound and unbound soliton pairs, as well as single solitons, can form in such interactions.
  If the interval between two incident beams is large relative to the width of their first lobes,
  the generated soliton pairs just propagate individually and do not interact.
  However, if the interval is comparable to the widths of the maximum lobes, the pairs interact and display varied behavior.
  In the in-phase case, they attract each other and exhibit stable bound, oscillating, and unbound states,
  after shedding some radiation initially.
  In the out-of-phase case, they repel each other and after an initial interaction,
  fly away as individual solitons. While the incident beams display acceleration, the solitons or soliton pairs
  generated from those beams do not.
\end{abstract}


 ] 

\noindent
In recent years, self-accelerating nondiffracting optical beams have been extensively studied \cite{siviloglou_ol_2007,siviloglou_prl_2007,ellenbogen_np_2009,chong_np_2010,efremidis_ol_2010,
bandres_njp_2013,bandres_oe_2013,alonso_ol_2012,kaminer_oe_2012,kaminer_prl_2012,aleahmad_prl_2012}.
Special attention has been focused on Airy \cite{siviloglou_ol_2007,siviloglou_prl_2007}
and Bessel beams \cite{durnin_josaa_1987,bouchal_cjp_2003}.
Analyses have been mostly confined to linear media, for the reason of wanting to observe minimally diffracting beams in linear optics.
According to the linear Schr\"{o}dinger equation (SE), the beam or the wave packet in the form of Airy function evolves
practically without diffraction and accelerates along a parabolic trajectory
\cite{berry_ajp_1979,siviloglou_ol_2007,siviloglou_prl_2007,siviloglou_ol_2008, bandres_ol_2008, bandres_ol_2009}.

In comparison with the thorough investigation of dynamics of single accelerating beams and related properties,
interactions between Airy beams have not attracted much attention.
Even though radially symmetric Airy beams readily display self-focusing in a nonlinear (NL) medium
\cite{efremidis_ol_2010,efremidis_pra_2013,liu_ol_2013},
the interaction of two Airy beams with varying distance between them has not -- but should have -- been investigated more deeply.
Indeed, the dynamics of an Airy beam in a NL medium has already been reported
\cite{kaminer_prl_2011,lotti_pra_2011,fattal_oe_2011,rudnick_oe_2011,panagiotopoulos_pra_2012},
but how will two Airy beams or two nonlinear accelerating beams behave if they propagate simultaneously in a NL medium?
These questions are addressed in this Letter.
Thus, we investigate the dynamics of two interacting Airy beams or accelerating nonlinear beams along the propagation direction, in one transverse dimension.
We study the beams that counterpropagate in different NL media, in particular a Kerr and a saturable photorefractive medium.
We vary the distance between the beams and consider the beams in-phase, as well as out-of-phase.

In paraxial approximation, the  normalized equation for the evolution of a slowly-varying envelope $\psi$ of the beam's electric field
is of the NLSE form \cite{zhang_oe_2010}:
\begin{equation}
\label{eq1}
  i\frac{\partial \psi}{\partial z} +\frac{1}{2} \frac{\partial^2 \psi}{\partial x^2} +\delta n \psi=0,
\end{equation}
where $\delta n$ is the NL change in the index of refraction, and
$x$ and $z$ are the dimensionless transverse coordinate and the propagation distance,
measured in units of some typical transverse size $x_0$ and the corresponding Rayleigh length $kx_0^2$.
Without $\delta n$ in Eq. (\ref{eq1}),
the equation is just the linear SE, one of whose exact solutions is the well-known Airy wave \cite{berry_ajp_1979}
with the characteristic infinite oscillatory tail.
To make it finite-energy,
this solution is generalized into \cite{siviloglou_ol_2007,efremidis_ol_2010}
\begin{align}
\psi(x,z)= &{\rm Ai}(x-z^2/4+iaz)\exp[i(6a^2z-12iax +6iaz^2 \notag \\
\label{eq2}&+6xz-z^3)/12],
\end{align}
which contains an arbitrary real decay constant $a\ge0$.
This solution is generated from an initial condition
$
  {\psi(x)={\rm Ai} (x) \exp(ax)}
$ and represents a finite-energy Airy beam.
The solution of Eq. (\ref{eq1}) without $\delta n$ will not be affected if it is shifted along the transverse coordinate
and scaled in the amplitude.
Thus, a more general form of an incident Airy beam can be written as
\begin{equation}\label{eq3}
\psi(x)=A \cdot {\rm Ai} [\pm (x\mp B)] \exp[\pm a(x\mp B)],
\end{equation}
in which arbitrary real constants $A$ and $B$ are introduced; they stand for the amplitude and the interval factor.
In addition, a linear superposition of such incident beams
will also be a solution of the equation.
\begin{figure*}[htbp]
  \centering
  \includegraphics[width=\textwidth]{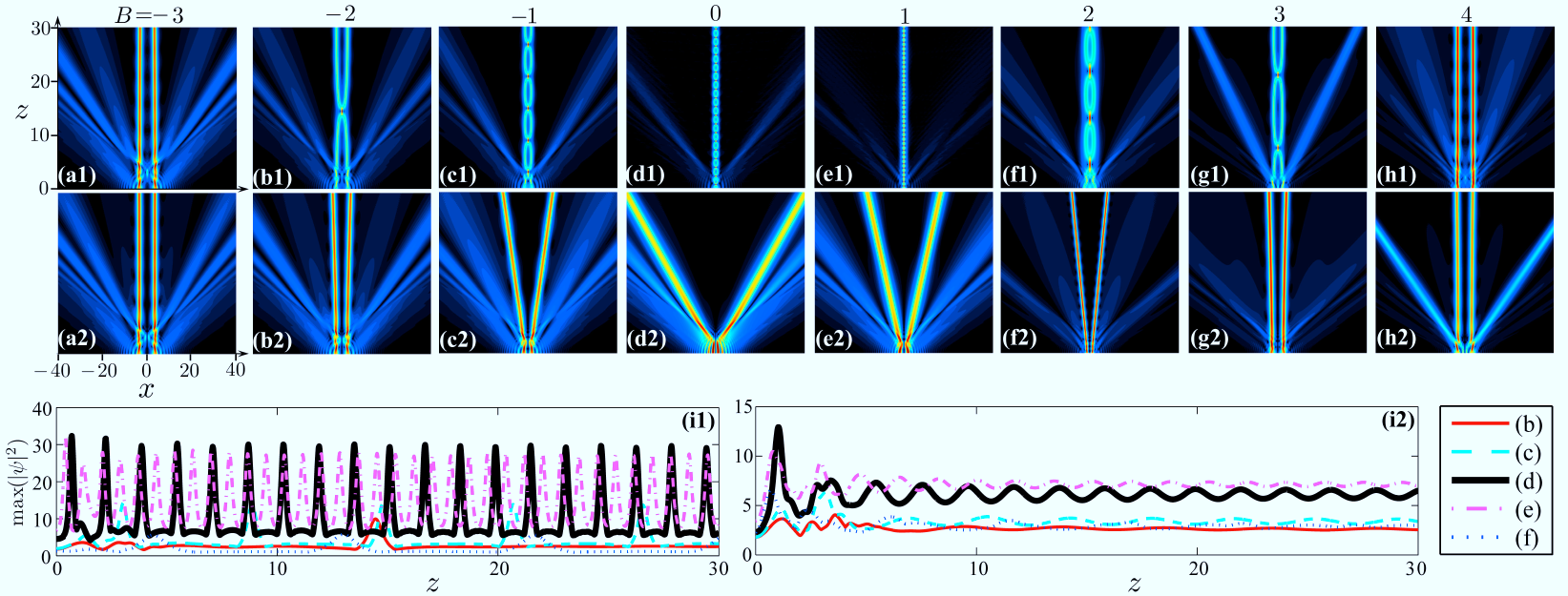}
  \caption{Soliton formation in the interaction of two in-phase ((a1)-(h1)) and out-of-phase ((a2)-(h2)) incident Airy beams, in Kerr medium.
  (i) Maximum intensities of beams in (b)-(f) versus propagation distance. }
\label{fig1}
\end{figure*}

We assume in this Letter that the incident beam is composed of two shifted counterpropagating Airy beams with a relative phase between them,
\begin{align}
  \psi(x)= & A \{{\rm Ai} [ (x - B)] \exp[ a(x - B)] + \notag \\
\label{eq4}  & \exp(il\pi) {\rm Ai} [- (x + B)] \exp[- a(x + B)]\},
\end{align}
where $l$ is the parameter controlling the phase shift.
If $l=0$, the two components are in-phase; if $l=1$, they are out-of-phase. Here, we restrict our attention to these two values.
Also, for simplicity, we take $a=0.2$ and $A=3$ throughout; of the three beam parameters present, we only vary $B$.
We now address different cases.

\textit{Kerr case.} We first consider the case with $\delta n =|\psi|^2$,
that is, the nonlinearity of the focusing Kerr-type.
We investigate the two cases of the incidence: in-phase and out-of-phase.
Since a large distance between Airy components in the incidence leads to a weak interaction,
we just show results with a relatively small distance.
The corresponding results are shown in Fig. \ref{fig1}.

Immediately visible is the considerable interaction and NL self- and mutual-focusing of the beams.
The major difference between the two cases is the attraction of beams in the in-phase case and the repulsion in the out-of-phase case.
For $B=-3$ and 4 in the in-phase case, the two Airy components form two parallel solitons after shedding some radiation,
as depicted in Figs. \ref{fig1}(a1) and \ref{fig1}(h1).
With the decreasing interval, the attraction between the two components increases
and bound breathing solitons are formed, with certain periods, as shown in Figs. \ref{fig1}(b1)-\ref{fig1}(g1).
The smaller the interval, the stronger the attraction
and the smaller the period of soliton breathing.

The maximum intensity versus the propagation distance, shown in Fig. \ref{fig1}(i1),
also demonstrates periodic properties of the soliton.
Curiously, the intensity image shown in Fig. \ref{fig1}(e1) has a smaller period than that in Fig. \ref{fig1}(d1),
even though $B=0$ in that case.
This can also be seen from the thick solid curve and the dash-dotted curve in Fig. \ref{fig1}(i1).
A smaller interval between beams should produce larger interaction, which should lead to a smaller period.
The reason is that the main lobe of the Airy beam with $B=0$ is located at about $-1$,
and there is still an interval between the two main lobes in the incidence.
So, the attraction is the biggest when $B=1$ and the period of the formed soliton is then the smallest.
It is also worth mentioning that the solitons are generated from the main lobes and
that the acceleration property of the main lobes is now absent \cite{fattal_oe_2011,kaminer_prl_2011}.
One can see in Fig. \ref{fig1} that solitons, as well as shed radiation, move along straight lines.

Concerning the out-of-phase case,
the results are shown in the bottom row in Figs. \ref{fig1}(a2)-\ref{fig1}(h2),
which shares the same numerical parameters as the in-phase case.
From the intensity images
one can see that the soliton pairs formed from the incidence actually repel each other,
and the maximum intensities for each case oscillate with propagation, as displayed in Fig. \ref{fig1}(i2).
The smaller the interval, the stronger the repulsion until the beams overlap.
However, when the beams strongly overlap, like in Fig. \ref{fig1}(e2), the repulsion decreases as the overlap increases.
Considering that the two Airy components are out-of-phase,
the main lobes will balance each other at $B=1$,
so that the distance between the secondary lobes (Fig. \ref{fig1}(e2))
is larger than the distance between the main lobes for $B=0$ (Fig. \ref{fig1}(d2)).
In other words, the soliton pair shown in Fig. \ref{fig1}(e2) is generated from the secondary lobes,
while the other are generated from the main lobes.
This is why the repulsion of the soliton pair in Fig. \ref{fig1}(d2) is stronger than that in Fig. \ref{fig1}(e2).
\begin{figure}[htbp]
  \centering
  \includegraphics[width=\columnwidth]{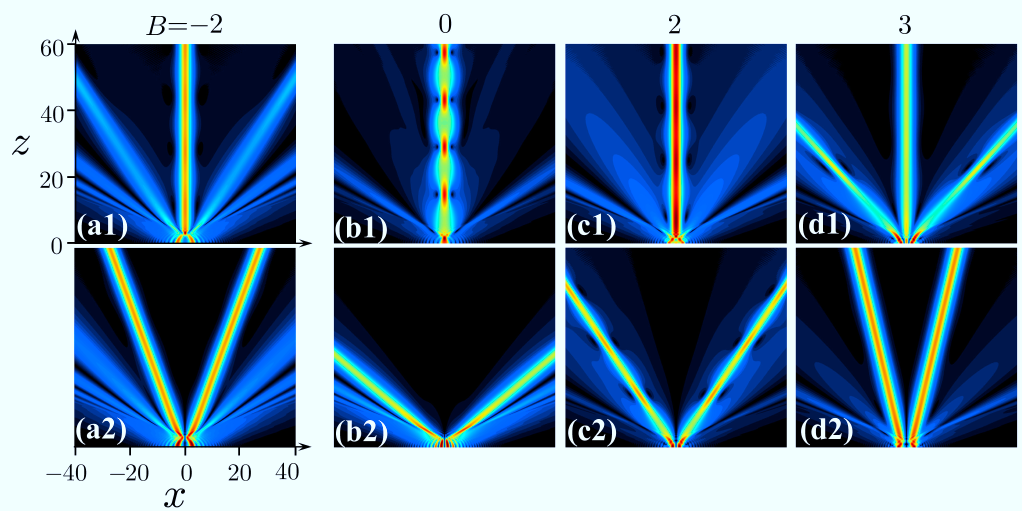}
  \caption{Intensities of two interacting in-phase (top row) and out-of-phase (bottom row) Airy beams in the
  medium with saturable nonlinearity.}
\label{fig2}
\end{figure}

It is interesting to note that in Fig. \ref{fig1}(h2), two soliton pairs are visible:
one pair comes from the main lobes of the Airy components and the other from the secondary lobes.
In all other figures only one pair is visible, in addition to the
excess radiation emanating initially from the interacting Airy beams.
Because the energy is mainly stored in the main lobes,
the intensity of the inner soliton pair is smaller than that of the outer pair,
but early in the propagation the two soliton pairs exchange energy at about $z=2$.
In addition, the repulsion of the outer soliton pair is stronger,
which comes from the main lobes possessing more energy.
We should note that these results will be different when $A$ is allowed to vary.
For small $A$ (less than 1), there will be no solitons generated;
for large $A$ ($\sim10$), multiple soliton pairs will be produced, but the propagation may become unstable.

\textit{Saturable case.} According to the standard theory of photorefractive effect,
we introduce the saturable nonlinearity in the form $\delta n =|\psi|^2/(r+|\psi|^2)$,
where $0 < r\le1$ is a saturation parameter that we take to be 1.
In Figs. \ref{fig2}(a1)-\ref{fig2}(d1) and \ref{fig2}(a2)-\ref{fig2}(d2)
we display the evolution of the incidence with two Airy components, in-phase and out-of-phase.

For the in-phase case, the two Airy components form individual solitons, as well as soliton pairs with breather-like behavior.
In the evolution shown in Fig. \ref{fig2}(a1),
the single soliton seen at $x=0$ is formed from the main lobes,
while the soliton pairs shown in the same figures come from the secondary lobes.
In Figs. \ref{fig2}(b1) and \ref{fig2}(c1), only single breathing solitons are formed; no soliton pairs are visible.
Since then the two Airy components are rather close,
the secondary lobes, together with the main lobes, contribute to the formation of the central solitons.
In Fig. \ref{fig2}(d1),
the solitons at $x=0$ are formed from the secondary and higher-order lobes,
while the soliton pairs come from the main lobes.

The out-of-phase case, as shown in Figs. \ref{fig2}(a2)-\ref{fig2}(d2),
is quite similar to the Kerr medium, in
that the superposition of two Airy components in the incidence
leads to the formation of repulsive soliton pairs. No individual solitons are formed.
But, if we compare Fig. \ref{fig2} with Fig. \ref{fig1},
we note that the repulsion between soliton pairs in the saturable NL medium is stronger than that in the Kerr medium.
We should mention that for small $A$, solitons or soliton pairs cannot form in the interaction;
however, the propagation in saturable NL medium is stable for arbitrary $A$, which is different from the Kerr medium.

\textit{Nonlinear accelerating beams case.} In addition to linear, there also exist
NL accelerating beams with parabolic trajectories; they are obtained
from Eq. (\ref{eq1}), when a traveling variable $x-z^2/4$
is introduced, to substitute for $x$
\cite{kaminer_prl_2011,lotti_pra_2011,panagiotopoulos_pra_2012}:
\begin{equation}
\label{eq5}
  i\frac{\partial \psi}{\partial z} - i\frac{z}{2} \frac{\partial \psi}{\partial x} +\frac{1}{2}
  \frac{\partial^2 \psi}{\partial x^2} +\delta n \psi=0.
\end{equation}
We seek NL accelerating self-trapped solutions of Eq. (\ref{eq5}) in the
form $\psi(x,z)=u(x)\exp{[i(xz/2+z^3/24)]}$ \cite{kaminer_prl_2011,lotti_pra_2011};
this leads to the equation
\begin{equation}
\label{eq6}
  \frac{\partial^2 u}{\partial x^2} +2\delta n u -x u=0.
\end{equation}
We treat Eq. (\ref{eq6}) as an initial value problem with the asymptotic behavior
$u(x)=\alpha\textrm{Ai}(x)$ and $u'(x)=\alpha\textrm{Ai}'(x)$ for large enough $x>0$;
here $\alpha$ indicates the strength of the nonlinearity induced by the potential solution.

\begin{figure}[htbp]
  \centering
  \includegraphics[width=\columnwidth]{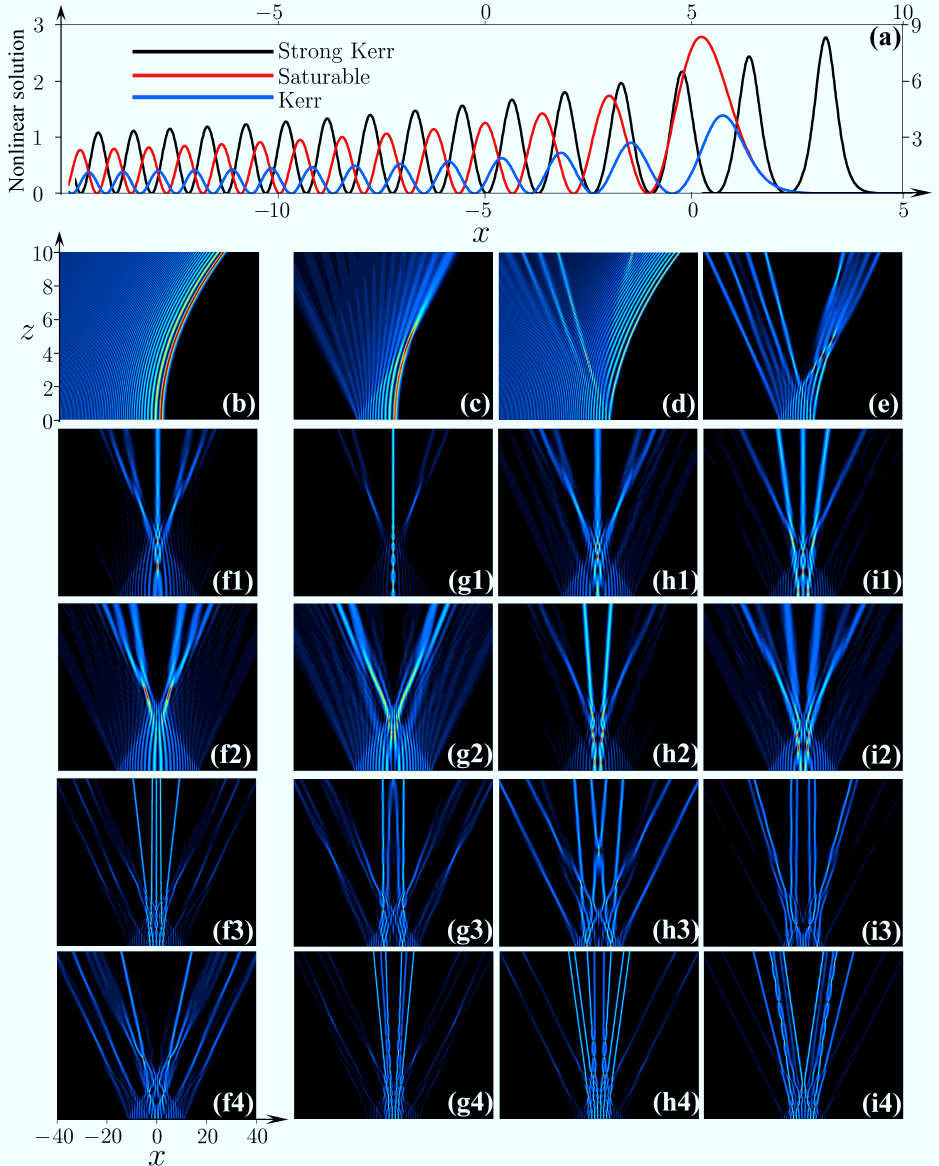}
  \caption{(a) Nonlinear accelerating modes, according to Eq. (\ref{eq6}). Kerr and saturable
  cases share the left and bottom axes, while the strong Kerr case uses the top and right ones.
  (b) and (c) Kerr and truncated Kerr solutions, obtained for $\alpha=10$.
  (d) and (e) Strong Kerr and truncated strong Kerr solutions, obtained for $\alpha=10^8$.
  (f1)-(i1) Interacting in-phase truncated Kerr solutions, shifted by $B=-$2, $-$1, 0, and 1,
  respectively. (f2)-(i2) Same as (f1)-(i1), but out-of-phase.
  (f3)-(i3) Same as (f1)-(i1), but for the in-phase truncated strong Kerr solutions.
  (f4)-(i4) Same as (f3)-(i3), but out-of-phase.}
\label{fig3}
\end{figure}

In Fig. \ref{fig3}(a) we display numerically obtained Kerr, saturable, and strong Kerr nonlinear accelerating modes.
Similar to Airy modes, they exhibit long tails and possess infinite energy.
As seen in Figs. \ref{fig3}(b)-\ref{fig3}(e), Kerr nonlinear solutions accelerate along parabolic trajectories.
Because of the infinite power,
it is reasonable to cut off oscillation tails and study the truncated cases, shown in Figs. \ref{fig3}(c) and \ref{fig3}(e).
One can see that the truncated ``normal'' Kerr solutions (obtained for $\alpha=10$) as shown in Fig. \ref{fig3}(c), shed radiation but
form no solitons. However, the strong Kerr solutions (obtained for $\alpha=10^8$), shown in Figs. \ref{fig3}(d) and \ref{fig3}(e),
readily form solitons from the strong radiation shedding.

Similar to the linear Airy beams from Eq. (\ref{eq4}), we now study the interaction
of two truncated nonlinear accelerating beams with the opposite propagation directions.
Figures \ref{fig3}(f1)-\ref{fig3}(i1) show the interaction of two in-phase Kerr beams
and Figs. \ref{fig3}(f2)-\ref{fig3}(i2) show the interaction of two out-of-phase beams.
Figures \ref{fig3}(f3)-\ref{fig3}(i3) and \ref{fig3}(f2)-\ref{fig3}(i2)
repeat the situation for the two truncated strong Kerr solutions.
It is seen that the accelerating beams easily generate single solitons and soliton pairs, which do not accelerate.
In comparison with the normal Kerr solutions, more soliton pairs can be induced from the strong Kerr solutions.
At the same time, solitons or soliton pairs cannot be formed if the intensity of the nonlinear accelerating beams is too small.

The behavior of the saturable nonlinear accelerating modes is similar to that of the Kerr nonlinear modes,
except that the saturable NL medium endorses stable propagation of beams with arbitrary high intensities.

In summary, we have demonstrated that
soliton pairs and even single solitons can be produced in the interaction of both in-phase and
out-of-phase Airy beams and the nonlinear accelerating beams in Kerr and saturable NL media.
Since the production of Airy beams \cite{siviloglou_prl_2007, ellenbogen_np_2009, cottrell_ol_2009, vo_josaa_2010, zhang_ol_2012}
and nonlinear accelerating beams \cite{lotti_pra_2011,dolev_prl_2012} has been reported in experiments,
an experimental observation of interactions reported in this Letter should be feasible.
Still, a number of interesting questions remains open,
such as: Can accelerating beams emerge from the interactions described in this paper?
Are these interactions elastic?
Can accelerating beams exchange momentum and energy during interactions?
How to describe interactions between accelerating beams and solitons in general?
We plan to address them in our future work.

This work was supported by
the China Postdoctoral Science Foundation (No. 2012M521773),
the National Natural Science Foundation of China (No. 61308015),
and the Qatar National Research Fund NPRP 09-462-1-074 project.
The 973 Program (No. 2012CB921804),
the National Natural Science Foundation of China (Nos. 61078002, 61078020, 11104214, 61108017, 11104216, 61205112),
the Specialized Research Fund for the Doctoral Program of Higher Education of China (Nos. 20110201110006, 20110201120005, 20100201120031),
and the Fundamental Research Funds for the Central Universities (Nos. xjj2013089, 2012jdhz05, 2011jdhz07, xjj2011083, xjj2011084, xjj2012080)
are also acknowledged.


\end{document}